# On a necessary criterion for stability of steady solutions of complex Ginzburg-Landau equation: a counterexample to the 'maximum entropy production principle'

Andrea Di Vita – DICCA, Univ. Genova, Via Montallegro 1, 16100, Genova, Italia

**Abstract**. *A maximum entropy production principle (MEPP) has been postulated to be a criterion of stability for steady states of open systems [Martyushev et al., Phys. Rep. **426**, 1 (2006)]. We find a necessary condition for stability of steady solutions of the complex Ginzburg-Landau equation. This condition violates MEPP.*

**§1. The problem.** It is customary to write the differential dS of entropy S in the form [1]

$$dS = d_i S + d_e S \qquad (1)$$

when considering a physical system which is far from thermodynamic equilibrium and is 'open' (i.e. it exchanges mass and energy with the external world). Here $d_iS$ and $d_eS$ are the contributions of irreversible processes occurring inside the bulk of the system and at the system boundary respectively. In contrast with 'closed' (i.e. not open) systems, where S = max. at thermodynamical equilibrium, the relative role of $d_iS$ and $d_eS$ in open systems far from thermodynamical equilibrium is far from being assessed even in the neighbourhood of steady states ($\partial / \partial t = 0$) -but for very restricted classes of problems [2].

In particular, spontaneous evolution towards a final steady state (henceforth referred to as 'relaxation') is related in closed systems to a growth of S towards its maximum value; thermodynamical equilibrium is the final state and the extremum property S = max. is related to stability of this state against perturbations. Here and below, we refer to final, steady, stable states as to 'relaxed states'; moreover, fluctuations occurring on spatial scales and time-scales much smaller than the characteristic spatial scales and time-scales of relaxation respectively are supposed to be averaged out when dealing with macroscopic quantities.

In contrast, in open systems boundary conditions may keep the relaxed state (if any exists) far from thermodynamic equilibrium. In the general case of continuous systems, it is customary [1] to write $d_iS = dt \cdot P$ and $d_eS = dt \cdot I$. Here we have introduced the quantities:

$$P \equiv \int_{bulk} dv \, \sigma_{bulk} \qquad ; \qquad I \equiv -\oint_{boundary} d\mathbf{a} \cdot \mathbf{j}_s \, dv, d\mathbf{a},$$

where $P$, $I$, dv, d**a**, $\sigma_{bulk}$ and $\mathbf{j}_s$ are the amount of entropy produced per unit time inside the system, the amount of entropy coming from the external world into the system across the boundary per unit time, the bulk volume element, the surface vector (pointing outwards) on the boundary, a scalar quantity representing irreversible processes within the bulk and a vector quantity which is related to the fluxes of matter and energy across the boundary respectively. In the following, we refer to dS/dt and $P$ as to the 'entropy production' and the 'bulk entropy production' respectively. The question if relaxed states in open systems enjoy extremum properties is still open.

It has been rigorously proven that relaxed states in open systems satisfy extremum properties (Prigogine's 'minimum entropy production principle', Onsager and Machlup's 'least dissipation principle') only if some very restrictive assumptions are satisfied, like Onsager symmetry relationships. (See e.g. [47] for a discussion of problems in fluid dynamics where minimum entropy production principle does not apply). The actual extremum property of interest depends on the particular boundary conditions [3]. Generally speaking, relaxation is described as the evolution of the system which starts from some initial condition inside a suitable attraction basin and ends at the configuration corresponding to the extremum quoted above.

If such restrictions are dropped, then a 'Maximum Entropy Production Principle' ('MEPP', i.e. $dS/dt$ = max.) in relaxed states has been postulated by many authors [4][5][6][7][8]. To date, attempts to derive MEPP rigorously meet no general consensus as they "are so far unconvincing since they often require introduction of additional hypotheses, which by themselves are less evident than the proved statement" [4] –see Appendix A.

Remarkably, a derivation of MEPP from the Second Principle of thermodynamics actually exists, but only provided that "the maximum thermodynamic flow is taken as a zero flow […] in practice, this can be realized, e.g. by time/space scaling" [23]. Generally speaking, then, a different choice of "time/space scaling" seems to lead to violation of MEPP, i.e. MEPP is not invariant under a ('scaling') transformation $\mathbf{x} \rightarrow \mathbf{x}' = a\,\mathbf{x}$, with $a > 0$. In contrast, some authors [24][25][26][27] suggest scale-invariance as a requirement for meaningful extremum properties. In the author's opinion, indeed, the latter properties should be invariant with respect to scaling transformations, like e.g. those involved when changing the system of units from SI to CGS.

Moreover, the results of [2] cast doubt on the very existence of a MEPP-like, general-purpose variational property of $dS/dt$ in relaxed states far from thermodynamic equilibrium; for selected classes of problems, however, validity of variational principles concerning either *I* or *P* separately remain possible. We show in Appendix A that many alleged applications of MEPP are indeed statements concerning *I*. The present work deals with the connection of scale-invariance with *P*.

In this paper, we are going to discuss in Sec. 2, 3 and 4 scale-invariance of *P* in three distinct classes of physical systems in steady state far from thermodynamic equilibrium, referred to as *A*, *B* and *C* respectively in the following. Classes *A*, *B* and *C* include the solutions of a simple kinetic equation [27], a class of compressible, electrically conducting fluids [28] and the solutions of complex Ginzburg-Landau equation (CGLE) [29] respectively; CGLE describes a wide class of weakly nonlinear phenomena. In particular, we are going to show how the systems of these classes map on each other via a scaling transformation [27] which conserves the value of *P* regardless of the detailed entropy-raising mechanism in each system. In Sec. 5 we invoke this scale-invariance of *P* and show that a necessary condition *P* = min. for stability of the steady state of a particular physical system of class *B* found in [36] in contrast with MEPP extends also to the steady states of corresponding systems in *A* and *C*. Conclusions are drawn in Sec. 6.

**§2. Class *A***. Let us focus our attention on a system *A* which is in contact with an environment at temperature T and where a Markovian process occurs which satisfies the Fokker-Planck equation:

$$\frac{\partial p}{\partial t} + \nabla \cdot \mathbf{b} = 0 \quad ; \quad \mathbf{b} = p\mathbf{a} - T\nabla p \tag{2}$$

Here p, **a** and **b** are a distribution function (normalized to 1), a rotational field (which plays the role of a thermodynamical force) and a probability current density respectively. Both **a**, **b** and p depend on time t and the coordinates **x**. Here and in the following, all quantities are dimensionless. Moreover, we limit ourselves to the case where $\mathbf{x} = (x,y,z)$ is the usual, 3D space vector so that $p(\mathbf{x})d\mathbf{x}$ is the probability of finding a particle in a small volume $d\mathbf{x}$ centered at **x**.

According to [27], a large class of scaling transformations exists, which leave both (2) and the value of *P* unaffected, regardless of the detailed mechanism of entropy production (i.e. of the physical nature of **a**). Physically, this invariance reflects invariance of thermodynamics with respect to changes of coordinates. Unlike *I*, *P* plays a central role in the models [42][43] based on Schnakenberg's treatment of master equation [44], which [26][27] are based upon; however, here and in the following our results rely in no way on the arguments of [42][43][44].

In the following, we focus on a sub-set of the transformations discussed in [27], namely on those transformations whose Jacobian $\Lambda$ does not depend on **x**:

$$\mathbf{x} \to \mathbf{x}' = a\mathbf{x} \quad ; \qquad p \to p' = p\, a^{-3} \quad ; \qquad \mathbf{a} \to \mathbf{a}' = a^{-1} \cdot \mathbf{a} \tag{3}$$

Here $a \equiv |\Lambda|^{1/3}$ is a constant quantity > 0 (see equation (10) of [27] for details). We are going to invoke (3) again and again, with various values of *a*, and write $\nabla' \equiv \partial/\partial\mathbf{x}' = a^{-1}\nabla$, etc.

**§3. Class *B***. Let us consider a system *B* of N charged particles with density n = Np, electric charge q and mass m. (The exact nature of these charged particles is not relevant here. Steady states may be achieved e.g. after suitable neutralization of net electric charge, which in turn may be due to a background of particles with electric charge –q. In this case, we may suppose that these background particles are so massive that their motion is negligible, so that we will consider them no more below). We take $a = m^{1/4}|q|^{-1/2}$. In steady state (2) and (3) lead to:

$$\nabla' \cdot \mathbf{b}' = 0 \quad ; \quad \mathbf{b}' = p'\mathbf{a}' - T\nabla'p' \tag{4}$$

We define $\mathbf{j} \equiv -N\mathbf{b}'$. After multiplication of all terms by $-N$, (3) makes (4) to be equivalent to:

$$\nabla \cdot \mathbf{j} = 0 \quad ; \quad \mathbf{j} = -\frac{nq^2}{m}\mathbf{a} + \nabla\Phi \tag{5}$$

where ($\chi$ is an arbitrary constant quantity)

$$\Phi = \frac{q^2 T}{m} n + \chi \tag{6}$$

We may consider $\Phi$ as a free variable in (5), then say that (6) just selects that system $B$ which satisfies (5) and where $P$ has the same value it has in $A$. Until now, we have provided no further information of $B$. We are free to consider $B$ as a particular electrically conducting, compressible fluid with velocity $\mathbf{v} \equiv \mathbf{j} / (n \cdot m)$, mean mass density $\rho_0$ and speed of sound $(\kappa \rho_0)^{-1/2}$; we allow both $\rho_0$ and the 'compressibility' $\kappa$ to be functions of T. In this case [30], a solution of (5) is:

$$\mathbf{j} = \frac{\gamma}{2i}(\psi^* \nabla \psi - \psi \nabla \psi^*) + w\rho \mathbf{a} \qquad (7)$$

where $\gamma > 0$ is a constant quantity (to be discussed below), $w \equiv -q/m$, $\rho \equiv n \cdot m$ and the field $\psi(\mathbf{x}) \equiv \rho(\mathbf{x})^{1/2} \cdot \exp(i\Phi/\gamma)$ satisfies the equation:

$$-\frac{\gamma^2}{2}\left(\nabla - i\frac{w}{\gamma}\mathbf{a}\right)^2 \psi - \frac{1}{\kappa \rho_0}\psi + \frac{1}{\kappa \rho_0^2}|\psi|^2 \psi = 0 \qquad (8)$$

According to (7), $\mathbf{j}$ is similar to the current density in a superconductor [31]. This analogy induces us to identify $\psi$ with an order parameter, and to localize entropy production within thin filaments whose radial size never shrinks to zero because of $\gamma > 0$ (in a superconductor $\gamma = \hbar/m$). Here the actual values of N, q, m, $\kappa$, $\rho_0$ and $\gamma$ are not relevant: what really matters is the fact that $P$ is the same in the steady states of $A$ and $B$. This is not worrying, as $B$ acts just as a dummy, auxiliary step in our discussion, and will disappear at the end. This freedom in choosing $\gamma$ allows us to retrieve a physical realization of $B$ in Section 5.

Finally, let L be a typical length of the system. It will be useful below to rewrite (8) as follows:

$$-\frac{r^2}{2}\Delta_y \eta + ir(\mathbf{c} \cdot \nabla_y)\eta + \left[\frac{|\mathbf{c}|^2}{2} - 1 + i\frac{r\,s}{2}\right]\eta + |\eta|^2 \eta = 0 \qquad (9)$$

where $\eta \equiv \rho_0^{-1/2}\psi$, $r \equiv L^{-1} \cdot \gamma \cdot (\kappa \rho_0)^{1/2}$, $s \equiv \nabla_y \cdot \mathbf{c}$, $\mathbf{c} \equiv w \cdot \mathbf{a} \cdot (\kappa \rho_0)^{1/2}$, $\mathbf{y} \equiv \mathbf{x}/L$, $\nabla_y = L \cdot \nabla$ and $\Delta_y = L^2 \cdot \Delta$.

**§4. Class C.** Equation (6) puts a condition on $\Phi$ and breaks the local gauge symmetry of (8). Together with the definitions of n, $\rho$ and $\psi$, equations (5)-(6) lead to:

$$ir(\mathbf{c} \cdot \nabla_y)\eta = ig\eta + ih|\eta|^2 \eta \qquad (10)$$

where $g \equiv (2wT)^{-1}\gamma \kappa \mathbf{j} \cdot \mathbf{a}$ and $h \equiv (2wT)^{-1}\gamma|\mathbf{c}|^2$. When deriving (10) we neglected both $\psi \nabla \rho - 2\rho \nabla \psi \propto O(1 - \exp(2i\Phi/\gamma))$ and $\rho \nabla \psi - \rho_0 \nabla \psi \propto O(|\nabla \psi|^2)$ in comparison to other terms $\propto O(|\nabla \psi|^0, |\nabla \psi|^1)$. The former assumption makes sense provided that we choose $B$ with a sufficiently large value of $\gamma$. The latter assumption makes sense as $\psi$ is an order parameter.

In turn, equations (9)-(10) give:

$$\zeta + (1 + i\mu)\Delta_q \zeta + \lambda(1 + i\nu)|\zeta|^2 \zeta = 0 \qquad (11)$$

where we have defined the following quantities $\zeta \equiv |\xi|^{1/2}\eta$, $\mu \equiv \alpha^{-1}\beta$, $\lambda \equiv |\xi|^{-1}\xi = \pm 1$, $\nu \equiv (\alpha - h\beta)^{-1}(h\alpha + \beta)$, $\xi \equiv (\alpha^2 + \beta^2)^{-1}(h\beta - \alpha)$, $\alpha \equiv -|\mathbf{c}|^2/2 + 1$, $\beta \equiv rs/2 + g$ and the operator $\Delta_q \equiv (\alpha^2 + \beta^2)^{-1}(\alpha r^2/2)\Delta_y$ which is just the Laplacian on the reduced coordinate $\mathbf{q} \equiv \mathbf{y}(2\alpha^2 + 2\beta^2)^{1/2}/(\alpha^{1/2} r)$.
\

Now, let us choose another value for *a* and apply (3)– with $\zeta \propto n^{1/2} \cdot \exp(i\chi/\gamma)$ and n ∝ p – and transform *B* into another physical system *C* with the same *P*. Then, equation (11) remains unaffected (after suitable rescaling). This is true even after repeated applications of (3): $C \to C' \to C'' \to \ldots$, each step with a different value of *a*. Information on *B* is lost, as anticipated.

If $\lambda = -1$, then (11) is the steady-state version of the complex Ginzburg-Landau equation (CGLE) [29]. CGLE describes a wide class of weakly nonlinear phenomena which occur in spatially extended continuous media and which are invariant under a global change of gauge.

We show in Appendix B that both μ and ν may take arbitrary values, and that we may reasonably take $\nabla\mu = \nabla\nu = 0$. Arbitrariness of μ and ν allows (11) to include many particular cases, like the nonlinear Schroedinger equation (μ→∞, ν→∞) [29], the reaction-diffusion problems near a Hopf bifurcation [32], the Stuart-Landau equation (after division by (1 + iμ) and further rescaling) in the long-wavelength limit $\Delta_q \to 0$ [33][34] and the Stewartson-Stuart equation [35] ($\lambda = + 1$) for Pouiseuille flow in fluid dynamics.

**§5 Stability.** Any perturbation of a relaxed state relaxes back to zero after a while, so that the final value of *P* coincides with the initial value *P* relaxed state. Then, (3) maps relaxed states of *A*, *B*, *C*… on each other.

A particular case of *B* exists, where a criterion of stability for the relaxed state is formulated with the help of *P*. According to [36], filaments of electric current in the pinch of a Dense Plasma Focus (DPF) [28] are a particular physical realization of equations (5)-(8) describing *B*. This pinch is a non-degenerate, viscous, resistive, electrically conducting fluid. Its life-time is >> the decay time-scale (as predicted by resistive magnetohydrodynamics), so we are allowed to speak of a 'steady state'; moreover, γ is proportional to the turbulent (Bohm [37]) heat diffusion; finally, coefficient and radiative and turbulent transport flattens $\nabla T$, so that T is as uniform in *B* as it was in *A*. Joule and viscous heating are the only entropy-raising mechanisms within the pinch bulk. Relaxed states satisfy [36][38] both Kirchhoff's [39] and Kortweg and Helmholtz' [40] principles of minimization of Joule and viscous heating power respectively. Since $\nabla T = 0$, this implies:

$$P_{\text{relaxed state}} = \min. \tag{12}$$

This extremum property acts as a necessary condition of stability against quasi-static perturbations [2][36]. (For example, Joule heating power is a Ljapunov function for the relaxed state [41] in the low magnetic Reynolds number limit which is relevant to DPF pinch [38]). Accordingly, if a quasi-static perturbation tries to move the system away from the relaxed state, then this perturbation makes the system to cross a succession of steady states 1,2,… where $P = P_1, P = P_2,\ldots$ with

$$P_{\text{relaxed state}} < P_1 < P_2 < \ldots \tag{13}$$

Since (3) leaves the value of P unaffected in all steady states, it follows that (13) holds also in all systems *A*, *C*, *C'*, *C''*…obtained from *B* through (3). All the way around, if we want to discuss stability of a system in steady state described e.g. by CGLE, then as a matter of principle we may map it on a DPF pinch with the help of (3); stability is only possible if (12) holds, because (3) maps relaxed states on relaxed states.

This result relies on no detailed knowledge of the entropy-raising mechanism in *A* and *C*. Then, (12) is a necessary condition of stability for the steady state of a system described by (2) or by (11) –including CGLE– even if physical mechanisms other than Joule and viscous heating (e.g., chemical reactions) rule *P*.

Admittedly, temperature is far from uniform in many systems of class *C* (like e.g. described by CGLE), unlike the class-*B* system discussed above. All the same, we show in Appendix B that the arbitrariness of μ and ν quoted above allows (3) to map class-*C* systems with $\nabla T \neq 0$ on class-*B* systems with $\nabla T = 0$. Then, our arguments leading to (12) for class-*C* systems are unaffected.

Let us consider the case where the solution $\zeta$ of (11) is the amplitude of a standing wave. Then, a relaxed state may exhibit either no oscillation ($|\zeta| = 0$) or finite-amplitude oscillation ($|\zeta| > 0$). According to (12), each case corresponds to a local minimum of *P* with its own basin of attraction in the phase space; maxima of *P* correspond to unstable configurations –in contrast with MEPP.

The validity of this result is not limited to class-*C* systems. For example, numerical simulations [45] show that when changes of a control parameter drive a system described by a simplified Fokker-Planck equation across a supercritical Hopf bifurcation from large limit-cycle like oscillations to small erratic oscillations around a fixed point, then *P* peaks very much near the transition, and decreases away from it. In our language, the limit cycle (the fixed point) is equivalent to a far-from-transition, stable oscillation with constant amplitude ≠ 0 (= 0) once small-scale, fast fluctuations have been averaged out. As the system is driven from one stable state to the other one, firstly P increases, then decreases: this is consistent with our result that each stable state corresponds to a local minimum of *P*.

**§6 Conclusions.** Starting from recent results concerning the scale-invariance of Markovian processes [27] based on Schnakenberg's treatment of master equation [44], we have shown that a large class of transformations exists which map on each other the steady-state solutions of (*A*) a simple Fokker-Planck equation [27], (*B*) the equations of motion of a particular class of electrically conducting, compressible fluids [30], and (*C*) the complex Ginzburg-Landau equation (CGLE) [29]. Moreover, these transformations leave also the amount *P* of entropy produced per unit time in the bulk of the system unaffected. Then, we show that these transformations allow us to extend to all systems *A*, *B* and *C* the necessary condition *P* = min [36] for stability of steady states of a particular physical system of class *B*, namely the pinch of a Dense Plasma Focus [28]. These results seem to be in contrast with the so called 'maximum entropy production principle', which has been postulated [4] to describe stable steady states far from thermodynamical equilibrium. Preliminary numerical analysis [45] of a system of class *A* seems to confirm our results. Application of the latter to the analysis of stable solutions of CGLE in fluid dynamics is the topic of future work.

**Acknowledgments**
Fruitful discussions with Prof. M. Polettini (Univ. Luxembourg) and Dr. E. Cosatto (AEN, Genova) are warmly acknowledged.

**Appendix A.** MEPP in the form of Ziegler's postulate [5][6] has "its statistical substantiation only if the deviation from equilibrium is small" [4] and is criticized in [9]. Moreover, the authors of [10] criticize the MEPP-supporting results of [7], which rely on Jaynes' 'information thermodynamics' [11]. In turn, it has been observed [2][26] that the author of [11] provides the fundamental tenet of equiprobability of microstates with no physical basis similar to what is provided e.g. by Liouville's theorem in equilibrium thermodynamics: "he swept the dirt under the carpet" [26].

Admittedly, many recent publications discuss evidence supporting MEPP [8][12][13][14][15][16][17][18][19][46] –for a review, see [4]. Remarkably, however, all these works seem to identify a property of $I$, rather than $dS/dt$. In fact, equation (7) of [8] suggests maximisation of the amount per unit time of the entropy of the Universe due to exchange between the relaxed system and the external world – in our language, maximisation of $-I$ only, not of $dS/dt$. The same limitation seems to apply to many alleged applications of MEPP, including metallurgy [12], crystallization [13], crystal growth [14], solidification [15], atmospheric physics [16][17][18][19] and plasma physics [46]. For example, the growth of a crystal is obviously affected by what happens at its boundaries. Accordingly: a) the selection criterion discussed in [12] is related to the "conjecture that each lamella must grow in a direction which is perpendicular to the solidification front"; b) the results of [13] are related to the growth velocity via the "rate of entropy production per unit area", which is a surface quantity like $I$ and unlike $P$; c) in [14] a crystal grows and changes its surface so as to select a flow which maximizes the entropy production, $I$ being the only contribution to the entropy balance which depends on both a flow and the surface of the system. As for solidification, the relevant quantity is the "entropy generation term ($S_{gen}$) which depends on the solid-liquid region" [15], i.e. on an interfacial boundary. As for atmosphere, the only contribution to the entropy balance investigated in [16][17][18][19] is due to energy fluxes across boundaries (see equations (5) and (14) of [16], equations (1)-(2) of [17], Fig. 1 of [18] and equation (1) of [19]). As for plasma physics, what is actually maximized in equation 7 of [46] is the temperature inhomogeneity in a boundary layer, a quantity related to $-I$.

In contrast with MEPP, different extremum properties hold for different terms in the entropy balance of relaxed states according to the analyses of [2] and Sec. 3 of [22] for systems where the approximation of local thermodynamical equilibrium holds and systems described by a Fokker-Planck equation respectively. For example, an extremum property similar to what is proposed in [8], and focused on $I$, is suggested in [20] for a shock wave in the limit of large Mach number, where the volume integral $P$ is as vanishingly small as the volume occupied by the shock front [2]. Noteworthy, MEPP fails to agree with models of the stratified atmosphere including viscosity and vorticity, where $P > 0$ [21].

**Appendix B**. We focus on the DPF pinch of Sec. 5. The quantities $\mathbf{a}$, $\nabla \wedge \mathbf{a}$ and $|\mathbf{c}|$ play the role of magnetic vector potential, of magnetic field and of Mach number respectively. The quantities $\mu$ and $\nu$ depend on both $\alpha$, $\beta$ and h. As for $\alpha$, we suppose –with no loss of generality– that the motion of the charge carriers in $B$ is subsonic, i.e. $\alpha \approx 1$. As for $\beta$, we choose the gauge on $\mathbf{a}$ in such a way that $\nabla \beta = 0$, so that $\nabla \mu = 0$; both $\beta$ and $\mu$ may take arbitrary values. As for h, the scaling $|\nabla \wedge \mathbf{a}| \propto \ln(1/R)$ at a small distance R from the filament axis [31] makes $\mathbf{c}$ to be a regular function of r even as $R \to 0$, hence h too is a regular function of $\mathbf{x}$. Now, we may suppose –again with no loss of generality, and in agreement with the 'presence of many fluxons' hinted at in [30]– that $B$ contains a large number of filaments, and replace therefore h with its spatial average, so that $\nabla h = 0$. (Discussion of the possible dependence of h on $\mathbf{x}$ goes outside the limits of this work). Addition of a component of $\nabla \wedge \mathbf{a}$ which is parallel to the filament axis leaves our argument unaffected, as such component scales as R for $R \to 0$. Again, h may take arbitrary values, as we may choose arbitrary (non-zero) values of w while leaving our arguments unaffected. The same conclusions holds for $\mu$ and $\nu$. Finally, if $h\beta - 1 > 0 \,(< 0)$ the $\lambda = + 1 \,(-1)$.

We are still left to show that (3) may map $\nabla T \ne 0$ systems on $\nabla T = 0$ systems. We limit ourselves to the $\alpha \approx 1$ case discussed above, for simplicity. If $\alpha \approx 1$, then $s \approx 0$, $h \approx 0$ and $\mu \approx \nu \approx \beta \approx g \equiv (2wT)^{-1} \gamma \kappa \mathbf{j} \cdot \mathbf{a}$, so that $\nabla \mu = 0$ corresponds just to $\nabla(T^{-1} \kappa \mathbf{j} \cdot \mathbf{a}) = 0$. This condition on T($\mathbf{x}$) is weaker than just $\nabla T = 0$. Now, formally the dependence of $\kappa$ on T is arbitrary (physically, indeed, the detailed value of the speed of sound $(\kappa \rho_0)^{-1/2}$ is not relevant in a subsonic theory). Accordingly, as far as T is a regular function of $\mathbf{x}$ we may take the validity of $\nabla(T^{-1} \kappa \mathbf{j} \cdot \mathbf{a}) = 0$ as granted, we no matter what j($\mathbf{x}$) and a($\mathbf{x}$) are like, and our results above hold even if $\nabla T \ne 0$.